\documentclass[aps,prb,floatfix,twocolumn,showpacs]{revtex4}
\usepackage{graphicx}
\usepackage{dcolumn}
\usepackage{bm}

\begin{document}

\title{Anomalous enhancement of tetragonality in PbTiO$_3$ induced
by negative pressure}

\author{Silvia Tinte, Karin M. Rabe and David Vanderbilt}

\affiliation{Department of Physics and Astronomy, Rutgers University,
Piscataway, New Jersey 08854-8019, USA}

\date{June 7, 2003}

\begin{abstract}
Using a first-principles approach based on density-functional
theory, we find that a large tetragonal strain can be induced in
PbTiO$_3$ by application of a {\it negative} hydrostatic pressure.
The structural parameters and the dielectric and dynamical
properties are found to change abruptly near a crossover pressure,
displaying a ``kinky'' behavior suggestive of proximity to a phase
transition.  Analogous calculations for BaTiO$_3$ show that the same
effect is also present there, but at much higher negative pressure.  We
investigate this unexpected behavior of PbTiO$_3$ and discuss an
interpretation involving a phenomenological description
in terms of a reduced set of relevant degrees of freedom.
\end{abstract}


\maketitle

\section{INTRODUCTION}

Recent work has shown that single-crystal solid-solution
ferroelectric perovskites can have dramatically improved
electromechanical properties compared to conventional transducer
materials.\cite{park97,nohe01,nohe02}  Representative materials
include PbZn$_{1/3}$Nb$_{2/3}$O$_{3}$--PbTiO$_3$ and
PbMg$_{1/3}$Nb$_{2/3}$O$_{3}$--PbTiO$_3$, which have ultrahigh
piezoelectric coefficients and low dielectric loss.  
These materials have also been observed to exhibit electric-field
induced phase transformations to ``ultrahigh'' strain states.\cite{park97}

PbTiO$_3$ serves as the common
parent compound for this class of materials, and may be supposed
to play an important role in the observed behavior.
Since the discovery of ferroelectricity in perovskite oxides
in the 1950's, PbTiO$_3$ has been the focus of
extensive experimental and theoretical study. 
It has a single phase transition at $T_c = 766$~K
from a paraelectric cubic phase to a ferroelectric tetragonal phase, and
a $c/a$ of 1.06 at low temperature.
The structure and properties of PbTiO$_3$ have been widely studied 
using first-principles
calculations.\cite{coh92,gar96,wagh97,sagh98,sai02}  Nevertheless, we
report here a feature of the behavior of PbTiO$_3$ that
had not previously been noticed.  Our calculations show that
an enormous
tetragonal strain can be induced in PbTiO$_3$ by application of a {\it negative}
hydrostatic pressure. The structural parameters, such
as cell volume and atomic displacements, are found to change abruptly
near a crossover pressure, displaying a ``kinky'' behavior
suggestive of proximity to a phase transition.  Analogous
calculations for BaTiO$_3$ show that the same effect is also
present there, but at much higher negative pressure.

In this paper, we
investigate this unexpected behavior of PbTiO$_3$, and discuss its
interpretation using a phenomenological description
in terms of a reduced set of relevant degrees of freedom.  
We make the notion of proximity to a phase transition more precise
by demonstrating that small changes
in the parameters in this description can take the system through a triple
point, leading to first-order transition behavior. Although the
application of negative pressure is not feasible experimentally,
our theoretical study provides useful insights into the
structural instabilities of PbTiO$_3$, and may ultimately help
suggest other, more practical avenues leading to enhanced
tetragonality in PbTiO$_3$ and related compounds.

The paper is organized as follows.  Section II provides the
technical details of our first-principles calculations.
Sec.~III gives the results of our computations for the
tetragonal distortion, structural parameters, and lattice
dynamical properties as a function of applied negative pressure.
In Sec.~IV we introduce a phenomenological model that allows us
to identify the relevant degrees of freedom responsible for
this behavior, and we conclude with a brief discussion in
Sec.~V.

\section{Computational details}

All {\it ab initio} calculations were performed in the framework of
the Hohenberg-Kohn-Sham density-functional theory~(DFT) within the
local-density approximation~(LDA).  We use the {\sf ABINIT}
package,~\cite{abi} a plane-wave pseudopotential code that, in
addition to ground-state total-energy and force calculations, allows 
linear-response computations of phonon frequencies and Born effective charges.     
Our calculations use the Perdew-Wang~\cite{perd} parameterization of the
Ceperley-Alder~\cite{ceper} exchange-correlation, and
the extended norm-conserving pseudopotentials of
Teter.~\cite{teter} These pseudopotentials include the O $2s$ and
$2p$, the Ti $3s$, $3p$, $3d$ and $4s$, the Ba $5s$, $5p$ and $6s$,
and the Pb $5d$, $6s$ and $6p$ in the valence states.
We have used an energy cutoff of 60 Ha throughout.  The integrals
over the Brillouin zone have been replaced by a sum over a
6$\times$6$\times$6 {\bf k}-point mesh.  Convergence of the
relaxations requires the Hellmann-Feynman forces to be less than
0.003 eV/\AA.  We have computed the eigenvalues and eigenvectors
of zone-center force-constant matrices by using both
finite-difference (frozen-phonon) and linear-response approaches
(with typical displacements of $\pm 0.007$\,a.u.\ for the former),
finding excellent agreement between the two schemes.

\section{RESULTS}

\subsection{Structural response}

The observed crystal structure of ferroelectric 
PbTiO$_3$ has space group $P4mm$, with Pb in the (1a) Wyckoff position 
(0, 0, $\xi_1$), Ti in (1b)($\frac{1}{2}, \frac{1}{2}, \frac{1}{2}+\xi_2$), 
and O in (1b)($\frac{1}{2}, \frac{1}{2}, \xi_3$) and
(2c) ($\frac{1}{2}, 0, \frac{1}{2}+\xi_4$) (0, $\frac{1}{2}, \frac{1}{2}+\xi_4$). 
The free structural parameters are the $a$ lattice constant,
the $c/a$ ratio and the atomic displacements along $\hat z$, $\xi_i$
(expressed in units of $c$).
At zero pressure, our LDA calculation yields a $T$=0 equilibrium lattice
constant of $a=7.301$~a.u. and $c/a=1.037$, which are $\sim -1\%$
and $\sim -3\%$ less than the experimental values of 7.373~a.u.~and
1.065,~\cite{exp} respectively. 
The fully-relaxed internal coordinates
are $\xi_{Pb}=0.0579$, $\xi_{Ti}=0.0268$,
$\xi_{O_{1},O_{2}}=-0.0335$ and $\xi_{O_{3}}=-0.0177$, where O$_1$
and O$_2$ are the ``in-plane" oxygens (2c) and O$_3$ is the apical
(along $\hat z$ (1b)) oxygen of the Ti-centered oxygen octahedron. 
We use the convention that $\sum_k \xi_k =0$.
Throughout this work the reference state will be our
theoretical minimum-energy ideal cubic structure ($a_0=7.331$~a.u.),
in terms of which the strains are defined as
$\eta_{1}=(a-a_0)/a_0$ and $\eta_{3}=(c-a_0)/a_0$.

\begin{figure}
\begin{center}
\includegraphics[width=7.6cm,angle=0]{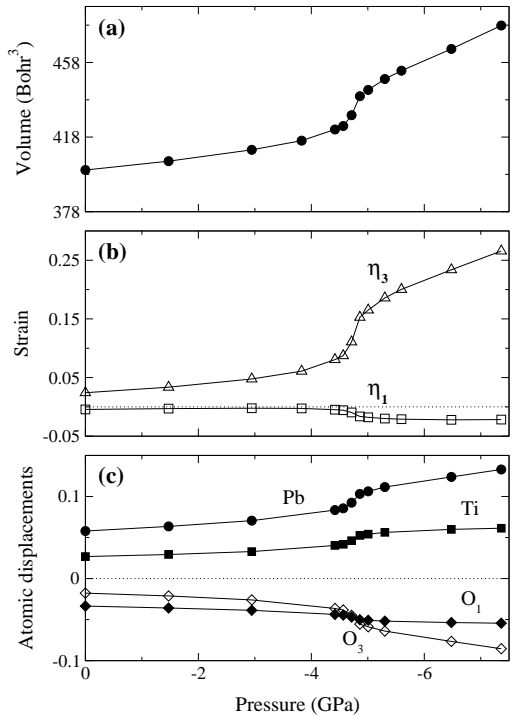}
\end{center}
\caption{Negative-pressure dependences of (a) volume (in Bohr$^3$);
(b) tetragonal strain $\eta_{3}$ and  $\eta_{1}$;
(c) atomic displacements in the $z$ direction (in $c$ units)
for tetragonal PbTiO$_3$.}
\label{fig:vol-p}
\end{figure}

We perform full, unconstrained optimization of the structural
parameters of tetragonal PbTiO$_3$ as a function of external
pressure ranging from 0 to $-$7~GPa. At a given pressure,
the set of parameters that minimizes the total energy is
determined.  The negative-pressure dependence of the optimized
structural parameters is shown in Fig.~\ref{fig:vol-p}, where
panels (a)-(c) display the unit-cell volume, strains ($\eta_{1}$ and
$\eta_{3}$), and internal atomic displacements in the $z$~direction,
respectively.  
As can be seen from the figure, all of the structural
parameters display an abrupt change around a crossover
pressure $p_c\simeq-4.8$~GPa in a way that suggests
proximity to a phase transition.  The rapid enlargement of the
unit-cell volume results from an abrupt increase of $\eta_3$, while it should
be noted that
the in-plane strain $\eta_{1}$ $\it decreases$ slightly in
the same pressure range.  The net effect of the negative pressure
is to stretch the unit cell strongly along $\hat z$ and
slightly squeeze it in the plane.  For example,
just above the transition at $p=-5.3$~GPa, we find
$\eta_{1}=-0.02$ and $\eta_{3}=0.186$, and the resulting $c/a$ is 1.21. 
Next, considering the relaxed atomic positions, the most
remarkable feature is the change in character of the
oxygen-displacement pattern. At pressures below $p_c$ the apical oxygen is
displaced less than the in-plane oxygens, whereas at pressures
above $p_c$ it is displaced more.  At $p_c$ both types of
oxygens are displaced by nearly the same amount, which means that the oxygen
cage forms a tetragonally strained octahedron.

\begin{figure}
\begin{center}
\includegraphics[width=6.6cm,angle=0]{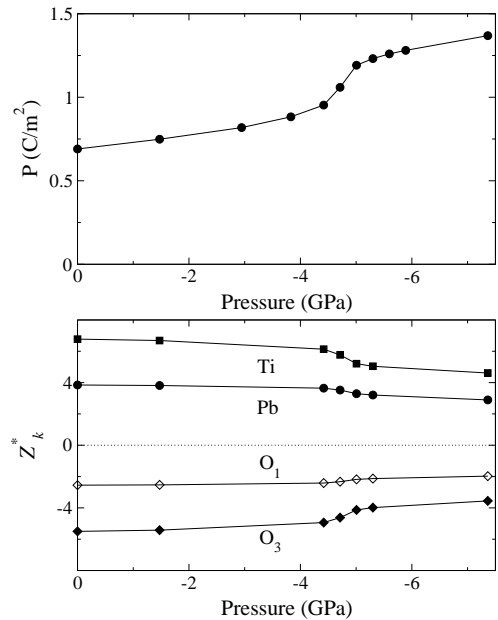}
\end{center}
\caption{Polarization (upper panel) and Born effective charges
($Z^{*}_k$) (bottom panel) for PbTiO$_3$
as a function of negative pressure.}
\label{fig:charge}
\end{figure}

For each of the optimized structures, we have also carried out
spontaneous polarization calculations (upper panel of Fig.~\ref{fig:charge})
as a function of negative pressure
using the Berry-phase theory of polarization.~\cite{king93}
It can be seen that the polarization shows the same kinky behavior
at $p_c$.  To get some insight into the polarization, we compute the Born
effective charges ($Z^{*}_k$).  The bottom panel of
Fig.~\ref{fig:charge} shows $Z^{*}_{k,33}$ in the relaxed
tetragonal structure at different pressures, as obtained by using
density-functional perturbation theory.  The results at zero pressure for $k =
$~Pb, Ti, O$_1$ and O$_3$ are 3.84, 6.77, $-$2.55, $-$5.50,
similar to previously calculated values of
3.92, 6.71, $-$2.56, $-$5.51,\cite{zho94} respectively.
However, a direct comparison is not possible because the latter
values were computed at the experimental tetragonal lattice
constants.  The most important features are the anomalously large
effective charges of Ti and O$_3$ (oxygen along the bond) compared
with their nominal charges ($+$4 and $-$2 respectively) and the
anisotropy of the oxygen charge.  These features persist throughout
the negative-pressure range of interest, although all of the
effective charges approach somewhat closer to the nominal values
above $p_c$.  Most notably, $Z^{*}_{Ti}$ and $Z^{*}_{O_3}$ change by
$\sim 25\%$ while passing through $p_c$, suggesting a weakening of
the Ti--O bond, whereas $Z^{*}_{Pb}$ and $Z^{*}_{O_1}$ decrease less
noticeably.  This follows the same trend observed in other
perovskites,~\cite{ghos98} where the $Z^{*}$'s decrease as the ions are
displaced away from their high-symmetry cubic sites.
The large changes in $Z^{*}_{Ti}$ and $Z^{*}_{O_3}$
reflect the change of the Ti environment along
the Ti--O chains; the shortened Ti--O$_3$ bonds remain almost
constant in length while passing through $p_c$, while the
elongated bonds lengthen abruptly there.  Note that
even though the $Z^{*}$'s decrease, the enhancement of the
ionic displacements is so strong that the overall effect is a
marked enhancement of the spontaneous polarization with
negative pressure.

\begin{figure}
\begin{center}
\includegraphics[width=8.cm,angle=0]{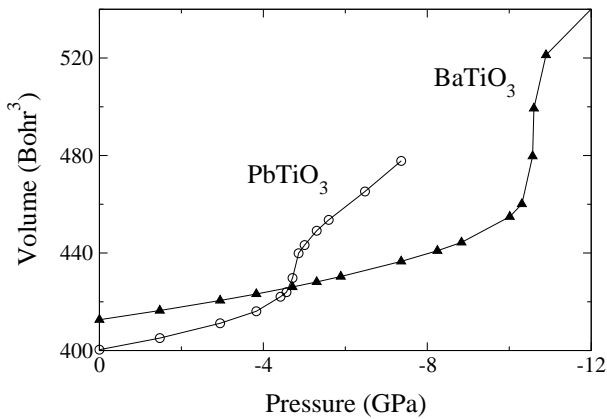}
\end{center}
\caption{Negative-pressure dependence of volume (in Bohr$^3$)
for tetragonal BaTiO$_3$ and PbTiO$_3$.}
\label{fig:bt}
\end{figure}

We have also investigated BaTiO$_3$ under negative pressure.  In
order to facilitate comparison with the results for PbTiO$_3$, we
have imposed tetragonal symmetry, even though the ground
state of BaTiO$_3$ is rhombohedral.  Interestingly, we find the
same anomalous effect as in PbTiO$_3$, but it occurs at much higher
pressure in BaTiO$_3$.  The computed optimized volume of BaTiO$_3$
in the tetragonal phase is plotted as a function of negative
pressure in Figure~\ref{fig:bt}, together with the corresponding
results for PbTiO$_3$ for comparison.  As seen, in BaTiO$_3$ the
jump of the volume occurs at $p_c\simeq -$10.6~GPa, a pressure that
is approximately twice as large as for PbTiO$_3$. Furthermore, the
abrupt volume enhancement is even larger; in a pressure interval of
width 1~GPa around $p_c$, the volume jump is $\sim 15\%$ in
BaTiO$_3$, to be compared with $\sim 5\%$ in PbTiO$_3$.

\subsection{Lattice instabilities}

As a first step towards understanding the unexpected behavior of
PbTiO$_3$ at negative pressures, we investigate the zone-center
lattice instabilities in a structure having the lattice constants
of the optimized tetragonal structure at the corresponding
pressure, but without internal distortions (i.e., all atoms at
centrosymmetric positions).  For each structure, we compute the
force-constant matrix at the $\Gamma$ point using the frozen-phonon method.

\begin{figure}
\begin{center}
\includegraphics[width=8.cm,angle=0]{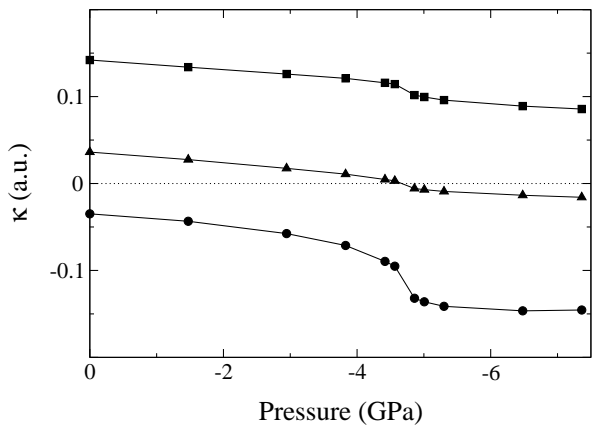}
\end{center}
\caption{Eigenvalues $\kappa$ of the force-constant matrix
in the centrosymmetric tetragonal structure with lattice
constants of the optimized tetragonal structure at the 
specified pressure.}
\label{fig:freq2}
\end{figure}

The tetragonal ferroelectric phase of PbTiO$_3$ belongs
to the $C^1_{4v}$ ($P4mm$) space group.
At the $\Gamma$ point, the vibrational representation is
spanned by two one-dimensional irreducible representations
$A_1$ and $B_1$ and one two-dimensional representation
$E$, of which there are 4, 1, and 5 copies, respectively, so
that the force-constant matrix is block diagonal.
As we are interested in the modes producing polarization along $\hat z$,
we only have to diagonalize the 4$\times$4 block of 
the $A_1$ subspace.  The pure translational mode
is discarded, and the eigenvalues ($\kappa$) of the three
remaining $A_1$ modes are plotted versus pressure in
Fig.~\ref{fig:freq2}.  
Negative values correspond to unstable modes. 
As expected, the ferroelectric
soft mode is already unstable at zero pressure, and it becomes
even more so as the crossover negative pressure region around
$p_{c}=-$4.8~GPa is crossed because of the cell-volume enhancement
that takes place there.  However, we also find that a second
mode becomes soft and that its frequency crosses through zero
at a pressure that is close to the same $p_{c}$.  At first sight,
this concurrence might be taken as a hint of some connection
between the second mode crossing and the anomalous ``kinky''
behaviors.
However, as will be explained in Sec.~IV, we think that if there
is such a connection, it is more likely that the rapid expansion
of the $c$ lattice constant causes the second mode crossing,
and not vice versa.

\begin{table}[b]
\caption {Comparison of eigenvalues ($\kappa$) and eigenvectors of
the original and second soft modes for different pressures.
Eigenvectors (Pb, Ti, O$_1$=O$_2$, O$_3$)
are normalized to unity.}
\begin{ruledtabular}
\begin{tabular}{lrrcrrrr}
 & $p$ (GPa)        &  $\kappa$ (a.u.) & $\phantom{a}$ &
     \multicolumn{4}{c}{Eigenvector}    \\
\hline
\multicolumn{2}{l}{Original} & & & & & & \\
& $0$     & $-0.187$  & & 0.520 & 0.573  & $-$0.388  &  $-$0.317 \\
& $-$4.56 & $-0.308$  & & 0.217 & 0.752 & $-$0.211  & $-$0.546  \\
& $-$6.48 & $-0.383$  & & 0.093 & 0.755 & $-$0.108 & $-$0.631 \\
Second &&  & & & & & \\
& $0$     & $0.190$ & & 0.713 & $-$0.682 & $-$0.076  & 0.121  \\
& $-$4.56 & $0.056$ & & 0.824  & $-$0.321 & $-$0.317  & 0.130  \\
& $-$6.48 & $-0.116$ & & 0.814  & $-$0.128& $-$0.394  & 0.102  \\
\end{tabular}
\end{ruledtabular}
\label{tab:mod-st}
\end{table}

The corresponding eigenvectors are also very sensitive to
pressure changes.
Table~\ref{tab:mod-st} summarizes the eigenvalues $\kappa$ and the eigenvectors
of the original ferroelectric soft mode and the second soft mode at three different
external pressures: zero, $p_c$, and a pressure higher than $p_c$.
As can be seen, the original soft mode shows the displacement pattern that is
typical of ferroelectric perovskites, with the cations moving in
opposition to the oxygen octahedra.  
As the negative pressure increases, the character of the mode evolves
to one dominated by the long-short alternation of the bonds in the Ti-O 
bonds along $\hat z$.
On the other hand, the pattern of the second soft mode consists
mainly of the Pb--O$_3$ plane moving against the Ti--O$_1$ plane.
At zero pressure the opposing cation displacements are dominant;
negative pressure leads to an increase in the O$_{1,2}$ displacements
and a decrease in the Ti displacement.

\subsection{Applied biaxial stress }

\begin{figure}
\begin{center}
\includegraphics[width=8.cm,angle=0]{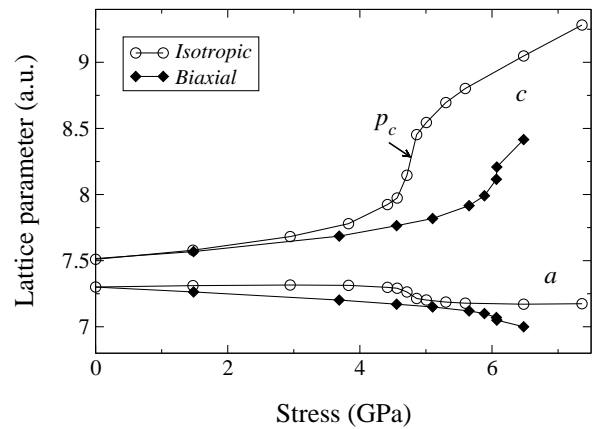}
\end{center}
\caption{Comparison of lattice parameters $c$ and $a$, in a.u.,
 as a function of absolute value of stress
 obtained for tetragonal PbTiO$_3$ by applying negative biaxial stress
 or positive isotropic stress (negative pressure).}
\label{fig:c-stress}
\end{figure}

As mentioned in Sec.~III\,A, the effect of negative isotropic
pressure on the cell shape is to stretch the unit cell along
$\hat z$ and slightly squeeze it in-plane.  Here we consider
what happens if instead we apply an in-plane stress only
($\sigma_{xx}$=$\sigma_{yy}$ and $\sigma_{zz}$=0).
This is effectively what happens in the case of epitaxial
growth on a substrate of slightly different lattice constant.\cite{per98}
In the calculation, we constrain the $a$ lattice parameter and allow
$c$ and all the ionic positions in the [001] direction to relax while
preserving the tetragonal space-group symmetry; the biaxial stress
needed to support this structure is then computed directly within
{\sf ABINIT}.

The results are shown in Fig.~\ref{fig:c-stress}, where the lattice
parameters are plotted for {\it expansive} (positive) isotropic
stress (negative pressure) and for {\it compressive} (negative)
in-plane stress.  This is the appropriate comparison since both have
the effect of enlarging the $c$ lattice constant and thus
potentially triggering the abrupt structural change.  Indeed, we
find that the biaxial stress {\it does} generate the same kind
of anomalous $c$-axis behavior.  The enlargement in $c$ occurs at
a somewhat larger magnitudes of the stress, and produces a somewhat more
modest $c$-axis expansion, than for the isotropic-pressure case.

It should be emphasized, however, that the abrupt variation only
appears when $c$ and $a$ are plotted vs.\ applied biaxial {\it
stress}; simply plotting $c$ vs.\ in-plane {\it strain} (i.e.,
vs.\ $a$) does not reveal anomalous behavior.  Moreover, in order to
obtain the same large $c$ values that result near the kink in the
negative-pressure case (indicated by $p_c$ in the figure), we find
that the in-plane lattice parameters would have to be compressed by
$\sim -3.2\%$, which is much larger than the typical compression
that can be attained by growth of PbTiO$_3$ on typical substrates
such as SrTiO$_3$ ($\sim -1.4\%$) \cite{chen95} or NdGaO$_3$ ($\sim
-1\%$).\cite{sun97}  Thus, it does not appear that the predicted
behavior can be observed in epitaxially strained films.  Perhaps
it may be possible to find some other way to generate a sufficiently
large biaxial compressive stress so that this effect can be observed.

In any case, we think it is interesting that the application of
compressive biaxial stress can generate the same kind of
anomalous structural behavior as the application of negative
isotropic pressure, at least for PbTiO$_3$.  This result tends
to support the speculation that some other kinds of variation
(e.g., chemical substitution) might also be capable of driving
a similar anomalous behavior.

\subsection{Applied electric field }

Another means of producing an effective tensile stress along $\hat z$
is the application of an electric field in the same direction. 
The induced
polarization should couple strongly to the tetragonal strain.
Therefore, we made a preliminary investigation of whether a strong
applied electric field might be capable of inducing a similar
anomalous enhancement of tetragonality.
The structural parameters in applied electric fields can be
estimated using an approximate scheme in which the {\it ab-initio}
force on each atom (computed in zero electric field) is augmented
by the product of the dynamical effective charge tensor for that
atom and the applied electric field vector.\cite{sai02,FuBellaiche}
Based on our preliminary results, but more specifically on the
results plotted in Fig.~8 of Ref.~\onlinecite{sai02}, it appears
that $\eta_3$ and $\eta_1$ do {\it not} show any abrupt change
up to $6 \times 10^3$~kV/cm.  Thus, it seems unlikely that the application of
an electric field can cause the same kind of anomalous behavior
as the application of negative pressure.  A more careful study
would be desirable, checking whether the above approximations might
be oversimplifying the treatment of some nonlinearities, but this
is left for a future investigation.

\section{Phenomenological model description}

In order to explore the origins of the anomalous behavior that has
emerged from our calculations in the vicinity of the crossover
negative pressure $p_c$, we turn now to an attempt to model this
behavior phenomenologically in terms of a reduced set of relevant
degrees of freedom.  We consider models similar to those that
underlie the effective-Hamiltonian scheme,\cite{zvr94,wagh97}
in which
the total energy is Taylor-expanded, in soft-mode and strain variables,
about a reference cubic phase.  The effect of the external hydrostatic
pressure $p$ is included by minimizing an enthalpy that includes
a $pV$ term in addition to the energy.

\begin{figure}
\begin{center}
\includegraphics[width=7.6cm,angle=0]{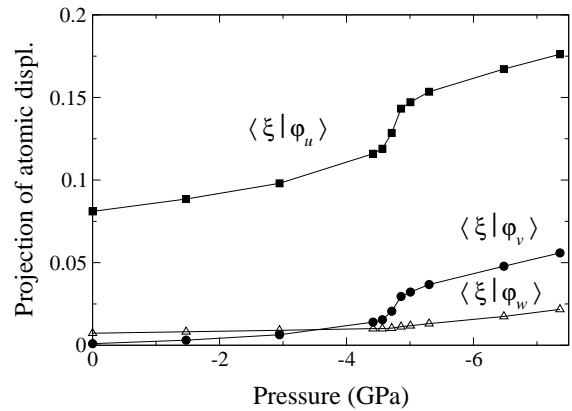}
\end{center}
\caption{Projection of the relaxed atomic displacements
along $\hat z$ at the specified pressure ($\langle\xi|$)
onto the three zone-center modes of the zero-pressure cubic structure
($|\varphi_u\rangle$, $|\varphi_v\rangle$ and $|\varphi_w\rangle$),
given in units of $c$.}
\label{fig:coef-p}
\end{figure}

The first step in the construction of the model is to determine
an appropriate set of degrees of freedom to be included.
For practical reasons, this set should be as small as possible, and
the definition of the degrees of freedom should not depend explicitly
on pressure.  Starting from a reference cubic structure, the fully
relaxed structure at a given pressure can be separated into two parts:
a homogeneous strain (leaving the atoms undistorted from their
high-symmetry positions) and an ``internal strain'' (i.e., internal
atomic displacements).  To describe the first part, we make the
simplest possible choice. As the unit-cell volume $V$ increases 
monotonically with negative pressure, it specifies the strain state
uniquely.
This choice has the added convenience that it 
is $V$ that directly couples to the pressure (it should be noted that 
the cell at given $V$ will in general be tetragonal and its value of $c$
and $a$ are thus functions of $V$).  
Next, we need to find an appropriate basis
able to capture the important internal strains at all relevant
pressures.  We first consider the entire set of zone-center modes of
the cubic structure at zero pressure, and then attempt to determine
which of these are most relevant for spanning the observed configurations
in the pressure range of interest.  Since the structure remains
tetragonal, the three non-trivial $A_1$ modes form the starting
point for this investigation.
Let $u$, $w$ and $v$ denote the
amplitudes of the softest, intermediate, and hardest mode
of the force-constant matrix,
with eigenvectors 
$|\varphi_u\rangle~=~|0.66, 0.39, -0.43, -0.43, -0.21\rangle$,
$|\varphi_w\rangle~=~|0.60, -0.79, 0.10, 0.10, -0.01\rangle$ and
$|\varphi_v\rangle~=~|0.06, 0.15, 0.33, 0.33, -0.87\rangle$, 
respectively. 
Then for each pressure, the optimized displacements shown in 
Fig.~\ref{fig:vol-p}(c) are projected onto these three modes 
to obtain the amplitudes plotted in Fig.~\ref{fig:coef-p}.
As can be seen from the figure, the largest contribution at all
pressures is clearly given by the soft mode $u$, while the other
two appear less important.  The contribution of the second mode
is almost independent of pressure and can therefore be neglected.
On the contrary, the highest-mode contribution is quite sensitive
to pressure, increasing notably at $p_c$. The need to include
at least one additional mode in the model subspace is clear from
Figure 1(c), as a single mode cannot capture the change in 
character above $p_c$. The figure suggests that the highest-mode
contribution $v$ is next in importance after the soft mode $u$; the
fact that including the $v$ mode yields
a qualitatively correct description of the distortions at all
relevant pressures will be confirmed below.

To develop the model, we consider the configuration subspace
defined by three independent variables: the unit-cell volume $V$,
the soft-mode amplitude $u$, and the hardest-mode amplitude $v$.
At each volume $V$, we start by considering only the mode $u$
(setting $v$=0), and Taylor-expand in $u$ as
\begin{equation}
E(V, u) = E_0 + c_0(V) +
c_{2u} \left(V\right) u^{2} + c_{4u} \left(V\right) u^{4} \;,
\label{eq:etot}
\end{equation}
where $E_0$ is the total energy at $V=V_0$ and $u=v=0$, so
that $c_0(V_0)=0$, and terms of order $u^6$ and
higher are dropped.
The coefficients $c_0$, $c_{2u}$ and $c_{4u}$ are obtained
by fitting to {\it ab-initio} total energies computed for
configurations corresponding to various values of $u$, and
the equilibrium value for $v=0$ and fixed $V$ is obtained by minimizing
Eq.~(\ref{eq:etot}), leading to
\begin{equation}
u^2_{eq} \left(V\right) = -\frac {1}{2} \frac {c_{2u} \left( V \right)}
{c_{4u} \left( V \right)}
\end{equation}
and
\begin{equation}
E(V, u_{eq}(V)) = E_0 + c_0\left(V\right)
- \frac {1}{4} \frac {c^2_{2u} \left(V \right)} {c_{4u} \left(V \right)} \;.
\label{eq:eumin}
\end{equation}
Then, starting at the configuration corresponding to $(u,v)=(u_{eq}(V),0)$,
we compute {\it ab-initio} total energies for
configurations corresponding to various values of $v$, holding $V$ and $u=u_{eq}(V)$ fixed, 
and fit
the result to quadratic order in $v$ as
\begin{equation}
E(V, u_{eq}(V),v)= E(V, u_{eq}(V)) + b_{1v}\left( V\right) v +
b_{2v}\left(V\right) v^{2} \;.
\end{equation}
Minimizing the total energy with respect to $v$ (at fixed $V$
and $u=u_{eq}$), the equilibrium amplitude of this mode is
obtained as
\begin{equation}
v_{eq} \left(V\right) = -\frac {1}{2} \frac {b_{1v} \left( V
\right)}
{b_{2v} \left( V \right)}
\end{equation}
and the total energy is
\begin{equation}
E = E_0 + c_0\left(V\right)
- \frac {1}{4} \frac {c^2_{2u} \left(V \right)} {c_{4u} \left(V \right)}
- \frac {1}{4} \frac {b^2_{1v} \left(V \right)} {b_{2v} \left(V \right)}
\;.
\label{eq:final}
\end{equation}

\begin{table}
\caption{Parameters of the model total-energy expansion.
All parameters are in atomic units. }
\begin{ruledtabular}
\begin{tabular}{l|cr|cr|cr}
V & $A_0$ & 4.4668 & $A_1$ & $-$0.0124 & $A_2$ & $-$5.69$\times$10$^{-5}$ \\
                 & $A_3$ & 2.37$\times$10$^{-7}$ & $A_4$ &
$-$2.17$\times$10$^{-10}$  & &  \\\hline
$u$ & $C_{2u}$ & 0.9264 & $C'_{2u}$ & $-$0.0042 & $C''_{2u}$ &
4.65$\times$10$^{-6}$\\
       & $C_{4u}$ & 0.7095 & $C'_{4u}$ & $-$0.0030 & $C''_{4u}$ &
3.18$\times$10$^{-6}$\\\hline
$v$ & $B_{1v}$ & 0.6700 & $B'_{1v}$ & $-$1.64$\times$10$^{-3}$ & & \\
       & $B_{2v}$ &$-$0.0851 & $B'_{2v}$ & 5.16$\times$10$^{-4}$ & &
\end{tabular}
\end{ruledtabular}
\label{tab:param}
\end{table}

To facilitate the modeling of the dependence of these results on
volume $V$, we fit the functions $c_0$, $c_{2u}$, 
$c_{4u}$, $b_{1v}$, and $b_{2v}$ as polynomials in $V$:
\begin{equation}
c_{0} \left( V \right) = A_0 + A_1 V +  A_2 V^2 + A_3 V^3 +  A_4 V^4 \;,
\end{equation}
\begin{equation}
c_{2u} \left( V \right) = C_{2u} + C'_{2u} V+ C''_{2u} V^2 \nonumber \;,
\end{equation}
\begin{equation}
c_{4u} \left( V \right) = C_{4u} + C'_{4u} V + C''_{4u} V^2 \;,
\end{equation}
\begin{equation}
b_{1v} \left( V\right) = B_{1v} + B'_{1v} V \;,
\end{equation}
\begin{equation}
b_{2v} \left( V\right) = B_{2v} + B'_{2v} V \;.
\end{equation}
The expansion parameters were determined by fitting to the results
of total-energy calculations in the interval of volumes from 400 to
480~a.u.$^3$, near $V_0=393.99$~a.u.$^3$.  The resulting parameters
of the model are reported in Table~\ref{tab:param}.

\begin{figure}
\begin{center}
\includegraphics[width=7.2cm,angle=0]{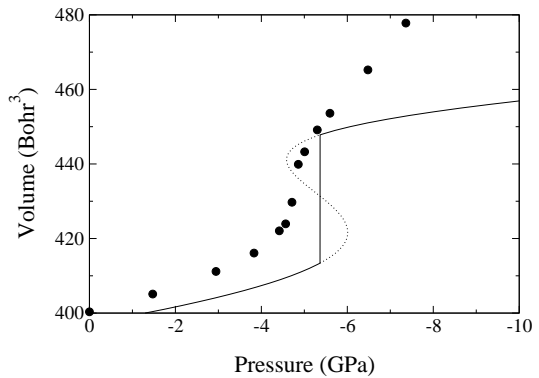}
\end{center}
\caption{Pressure computed at different volumes
by using the phenomenological model (full line)
compared with the first-principles results
(circles).}
\label{fig:model}
\end{figure}

We then explored the behavior of this model by determining
the equilibrium structural parameters as a function of 
pressure $p$, minimizing the enthalpy $H=E+pV$.  The result
is shown by the solid line in Fig.~\ref{fig:model}.
We find that there is a pressure interval from about
$-4.6$ to $-6$ GPa in which two local minima compete,
with a first-order isostructural transition between minima at
$p_c\simeq-5.4$ GPa.  The locations of the secondary minimum and
the saddle point are indicated by the dotted line in
Fig.~\ref{fig:model}, obtained by evaluating the
thermodynamic relation $p = - \partial E(V)/\partial V$
as a function of $V$.  

Comparing the behavior of the model with the first-principles
results, shown as solid circles in Fig.~\ref{fig:model},
we see that the model correctly reproduces the existence of
an abrupt variation of structural parameters with negative
pressure.  In fact, the model even goes too far, exhibiting
a true first-order transition where there is none in the
first-principles results.  The model also slightly
overestimates, by about 10-15\%, the magnitude of the negative pressure 
at which the abrupt change occurs, and underestimates
the pressure dependence of the volume at the highest negative
pressures.  

\begin{figure}
\begin{center}
\includegraphics[width=7.2cm,angle=0]{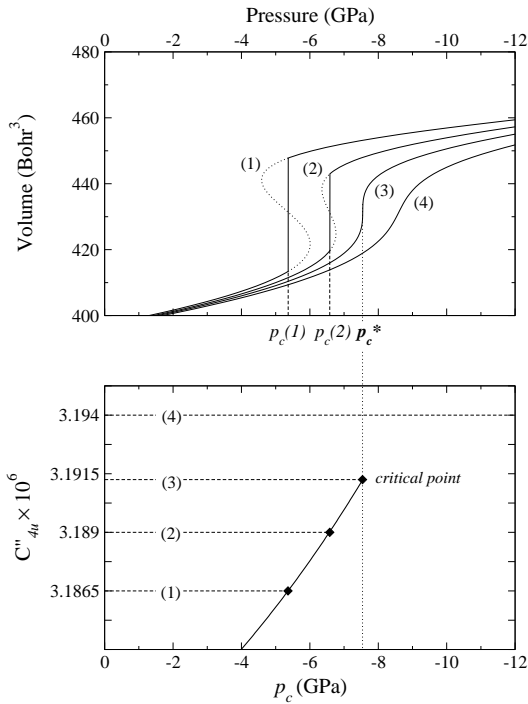}
\end{center}
\caption{Calculated $P-V$ phase diagram for different
$C''_{4u}$ parameters (upper panel).
The original model is marked as case (1).
$C''_{4u}$ parameter--critical pressure ($p_c$) diagram
(bottom panel).
For values bigger than $C''_{4u}$ at the critical point
there is no phase transition. }
\label{fig:tuning}
\end{figure}

Even with these minor discrepancies, the model is very useful
in generating a clearer picture of the observed behavior. In
particular,
the fact that a first-order transition occurs in the model
supports the idea that the anomalous enhancement of tetragonality in the
first-principles calculations results from proximity to a phase
transition.  To make this idea more precise, we consider small
variations of one of the model parameters, $C''_{4u}$, keeping all
other parameters fixed, and study the transition behavior as a function
of pressure.  The upper panel of Fig.~\ref{fig:tuning} shows the
$P$--$V$ diagram for four different values of $C''_{4u}$, starting
from the original model marked as case (1) and increasing $C''_{4u}$
incrementally to case (4).  As can be seen, for the smaller values
of $C''_{4u}$ there is a first-order phase transition from a
low-$c/a$ to a high-$c/a$ phase at a transition pressure $p_c$
(cases (1) and (2)).  The volume discontinuity decreases with
increasing $C''_{4u}$ and vanishes at a triple point (case (3)).
In other words, a critical value of the $C''_{4u}$ parameter exists
above which there is no distinction between two phases (see, e.g.,
case (4)).  The relation between $C''_{4u}$ and the transition
pressure is presented in the bottom panel of
Figure~\ref{fig:tuning}; the phase boundary defined in this way
terminates at the triple point identified by pressure $p_c^*$.

Thus, Fig.~\ref{fig:tuning} demonstrates that with increasing
negative pressure, PbTiO$_3$ passes very close to a triple point. In fact, 
it passes
so close that the errors introduced by the truncations
and simplifications of our model, described earlier in this
section, are sufficient to cause the model system to exhibit
a true first-order phase transition.
Indeed, to shift the model system into the transition-free
region of parameter space,
it suffices to change $C''_{4u}$ by only $\sim 0.15\%$. 
A more accurate model containing additional fitting
parameters (e.g., an independent treatment of lattice constants
$c$ and $a$, inclusion of mode $w$, and/or a more careful treatment
of cross terms between modes $u$ and $v$) would presumably be
sufficient to reproduce the observed ``kinky" behavior without
the spurious first-order transition.

\begin{figure}
\begin{center}
\includegraphics[width=7.2cm,angle=0]{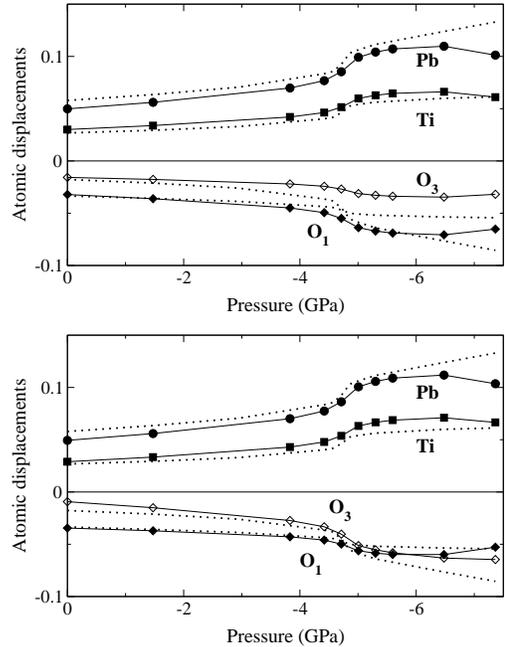}
\end{center}
\caption{Configurational space described by $u_{eq}$ (upper panel)
and by $u_{eq}$ and $v_{eq}$ (bottom panel)
compared with the minimum-energy structures (dotted lines).}
\label{fig:projec-u}
\end{figure}

We now turn to an examination of the role of the additional
mode $v$ in the anomalous negative-pressure behavior of PbTiO$_3$.
The effects of holding $v$=0 can be easily obtained by minimizing
the enthalpy $E+pV$ with $E$ from Eq.~(\ref{eq:eumin}).
The $V$ vs.\ $p$ curve still exhibits the anomalous negative-pressure behavior, 
although
the transition region is considerably shifted (to around $-12$ GPa)
and there is no first-order transition in this case (similar to the
first-principles result).  
Thus, inclusion of the additional mode $v$ is not absolutely
necessary to produce the anomalous negative-pressure behavior.
On the other hand, neglecting this mode leads to a substantial
quantitative error in the transition pressure. Moreover, as
we have already seen in Fig.~\ref{fig:vol-p}(c) and
Fig.~\ref{fig:coef-p}, the inclusion of the mode $v$ is needed to
correctly capture the variation of the atomic displacements with pressure.  
This point is emphasized in Fig.~\ref{fig:projec-u}.
In the upper panel,
the ratios of the displacements are fixed with an overall
amplitude proportional to $u$ (shown with symbols), 
so that it is impossible to
reproduce the crossing in the oxygen displacement pattern 
around $p_c$. In the lower panel, the inclusion of the $v$
mode yields a qualitative improvement, with a crossing 
close to $p_c$. The discrepancy between the
model and the first-principles results (shown with dotted lines) 
grows much more
noticeable above $p_c$. This most likely reflects the contribution
of the neglected modes and couplings to the strong pressure dependence above
$p_c$, which is underestimated by the model.

Finally, we analyze the relation between the modes of the centrosymmetric
tetragonal structure (dependent on pressure)
and the modes $u$, $v$ and $w$, used in the construction of our model.
In our discussion of Fig.~\ref{fig:freq2} in Sec.~III\,B, we noted
that the tetragonal mode of intermediate hardness crosses from positive to
negative $\kappa$ (i.e., from stable to unstable behavior) at a
pressure very close to that at which the anomalous structural
response occurs. 
Projecting the pressure-dependent tetragonal modes 
on the cubic modes, we find that
the tetragonal soft mode 
is mainly described by the mode $u$ at all pressures, with 
a significant contribution of the mode $w$ above $p_c$,
whereas the tetragonal mode of intermediate hardness
is mainly described at lower pressures by
the mode $w$, and then above $p_c$ by the mode $v$.
Since the kinky behavior occurs even when both 
cubic-structure modes $v$ and $w$ are frozen to zero amplitude,
we think that it is unlikely that
the zero crossing of the second tetragonal-structure mode has a
{\it causal} role in the anomalous structural behavior.  Instead,
it seems more likely that it simply {\it results} from the abrupt
expansion of the $c$-axis lattice parameter.

Thus, we believe that the model developed in this section, based on
an expansion of the total energy with respect to the volume and two
zone-center mode amplitudes, is a reasonable compromise between
simplicity and accuracy.  It gives a good qualitative and even
semi-quantitative description of the abrupt variation of the
structural parameters near $p_c$.  The fact that it predicts a
true first-order transition at $p_c$, instead of a smooth but ``kinky"
cross-over, should be taken more as a sign of the proximity of the
real system to a phase transition than as a failure of the model.
We are hopeful that this model will be useful for investigating
and describing similar behavior in other perovskite systems.

\section{DISCUSSION AND SUMMARY}

We have shown that the application of negative pressure
induces a large enhancement of tetragonal strain in tetragonal PbTiO$_3$. 
Specifically, in a window of pressure centered
around a particular value $p_c$, $c/a$ abruptly increases,
and all structural parameters exhibit a corresponding ``jump" 
that suggests proximity to a phase transition.

To describe this unexpected behavior in PbTiO$_3$,
we have generated a phenomenological model from our {\em ab~initio}
results, based on an expansion of the total energy with 
respect to a carefully selected subset of the structural degrees of freedom.
This has led to the identification of a phase boundary,
analogous to the liquid-gas phase boundary, in the model
parameter space.
With the parameters obtained from fitting to first-principles
total energies of PbTiO$_3$,
the model predicts not just a ``kinky" behavior, but a true 
first-order transition at a critical pressure
$p_c$ close to the pressure at which the anomaly occurs in
the first-principles calculations.
We found that a tiny variation of one parameter
is sufficient to drive the model through a triple point 
and bring it to a transition-free portion of parameter space
where the ``kinky" behavior is qualitatively well-reproduced.
It might become possible to realize such variations of effective
model parameters by some combination of ``external fields'' such
as chemical substitution, temperature, epitaxial stress, or
homogeneous electric field.  If so, one might develop experimental
techniques for turning the transition off or on, tuning
the associated transition properties, and perhaps even exploring
the region of the triple point itself.

Finally, we speculate that this ``anomalous" behavior 
of the structural parameters of tetragonal PbTiO$_3$ 
could be a more general feature in perovskite oxides.
For example,
we have found that the same anomalous behavior is also present
in BaTiO$_3$ with imposed tetragonal symmetry (though at much
higher negative pressure), so that a similar 
phase boundary could be explored in this portion of perovskite
parameter space. The possibility of analogous transitions in
other ABO$_3$ perovskites is under investigation.
This could present tantalizing opportunities for designing
perovskite-based materials with large and controllable strain
variations.

\begin{acknowledgments}

We would like to thank Na Sai for providing the initial suggestion for
the direction of this work.
This work was supported by the Center for Piezoelectrics by Design
(CPD) under ONR Grant N00014-01-1-0365.  Computational facilities for
the work were also provided by the CPD.
We thank Morrel Cohen for useful discussions.

\end{acknowledgments}

\end{document}